\documentclass[%
 prl,
 twocolumn,
superscriptaddress,
 amsmath,amssymb,
 aps,
]{revtex4-1}
\usepackage{amsfonts,amssymb,amsmath,mathrsfs}
\usepackage{graphicx}
\usepackage{dcolumn}
\usepackage{bm}
\usepackage{subfigure}
\usepackage{upgreek}
\usepackage{nicefrac}
\usepackage[usenames, svgnames]{xcolor}
\newcommand{\im}{\mathrm{i}}
\newcommand{\e}{\mathrm{e}}

\begin{document}

\preprint{APS/123-QED}

\title{Tailored dissipation for directional transport in plasmonic ratchets}

\author{Anna Sidorenko}
\thanks{Equal contribution}
 \email{a\textunderscore sidorenko@uni-bonn.de}
 \affiliation{Physikalisches Institut, Rheinische Friedrich-Wilhelms-Universität Bonn, Kreuzbergweg 24, 53115 Bonn, Germany}
 
\author{Jan Mathis Giesen}%
\thanks{Equal contribution}
 \email{m.giesen@rptu.de}
\affiliation{Physics Department and Research Center OPTIMAS, RPTU University Kaiserslautern-Landau, D-67663 Kaiserslautern, Germany}%


\author{Sebastian Eggert}
\affiliation{Physics Department and Research Center OPTIMAS, RPTU University Kaiserslautern-Landau, D-67663 Kaiserslautern, Germany}%

\author{Stefan Linden}
\affiliation{Physikalisches Institut, Rheinische Friedrich-Wilhelms-Universität Bonn, Kreuzbergweg 24, 53115 Bonn, Germany}%


\date{\today}

\begin{abstract}

We present a joint experimental and theoretical study of a ratchet implemented in arrays of evanescently coupled plasmonic waveguides with tailored losses. 
In this setup the time-periodic dissipation is the only active mechanism and notably, we find better rectified transport and lower losses in the 
transmitted signal with increased local dissipation.
Using Floquet theory, we uncover a driving regime that allows efficient directional transport for suitable driving frequencies and loss rates, which are linked to linear quasienergy bands with minimal losses. These regions are separated from non-resonant behavior by sharp transitions with characteristic exceptional points in the spectrum.
Direct experimental observation of the Floquet-dissipative ratchet effect using a combination of real- and Fourier-space leakage radiation microscopy is provided. 
\end{abstract}

\maketitle


{\it \label{sec:level1}Introduction.}
The ratchet effect refers to the appearance of a directed motion in the absence of a global biased force. The underlying principle of  ratchets is based on a non-equilibrium system with locally 
broken space- and time-reversal symmetries~\cite{PhysRevLett.84.2358, Denisov_2007, PhysRevA.75.063424,PRXQuantum.5.030356}. 
Ratchet phenomena have been observed across a wide range of systems, from molecular motors \cite{pumm2022dna} and colloidal particles \cite{leyva2022hydrodynamic} to cold atoms \cite{dupont2023hamiltonian, science1179546, PhysRevLett.94.164101}, microbiological \cite{Mahmud2009} and semiconductor devices \cite{custer2020ratcheting,kedem2017drive, science.286.5448.2314}, as well as photonic structures \cite{blessan2025directional,sanchez2019phase, FedorovaDirFilter,matsubara2022polarization}.

Two types of ratchets with periodic driving mechanisms are known \cite{DENISOV201477}: The first type is in the class of Brownian motors \cite{Rousselet1994,PhysRevLett.79.10,PhysRevLett.94.164101,RevModPhys.81.387,bahrova2024skew}, which 
rely on a tailored stochastic environment to achieve directed transport
in open non-equilibrium systems \cite{Wanjura2020, downing2020chiral, PhysRevLett.134.093801}, making dissipation a resource rather than a restriction~\cite{SURAJ2025131435, Fedorova2020, jurgensen2025quantized, janovitch2025active, hong2025observation,wzb3-dbg9}.
The second type is the class of Hamiltonian ratchets \cite{dupont2023hamiltonian,PhysRevA.75.063424,DENISOV201477,science1179546}, where
the directed transport is fully coherent without an environment. 
In this case, the initial condition is crucial in determining the direction of the current. 
This invites the question, if it is possible to combine the advantages of both types, namely using deterministic non-stochastic time-evolution, but 
with a robust transport direction independent of the initial condition.  We now propose non-Hermitian Hamiltonian time-evolution to construct such a third type of  
ratchet using only {\it time-periodic damping}, but 
without any applied forces and without any stochastic environments.
 A corresponding minimal model is presented in this Letter.
This ratchet is implemented
experimentally by tailored losses in arrays of coupled plasmonic waveguides.
Notably, we observe frequency windows with strong rectification, where the directed transport {\it increases} with larger local damping.

{\it Model.}
\begin{figure}
    \centering
    \includegraphics[width=0.4\textwidth]{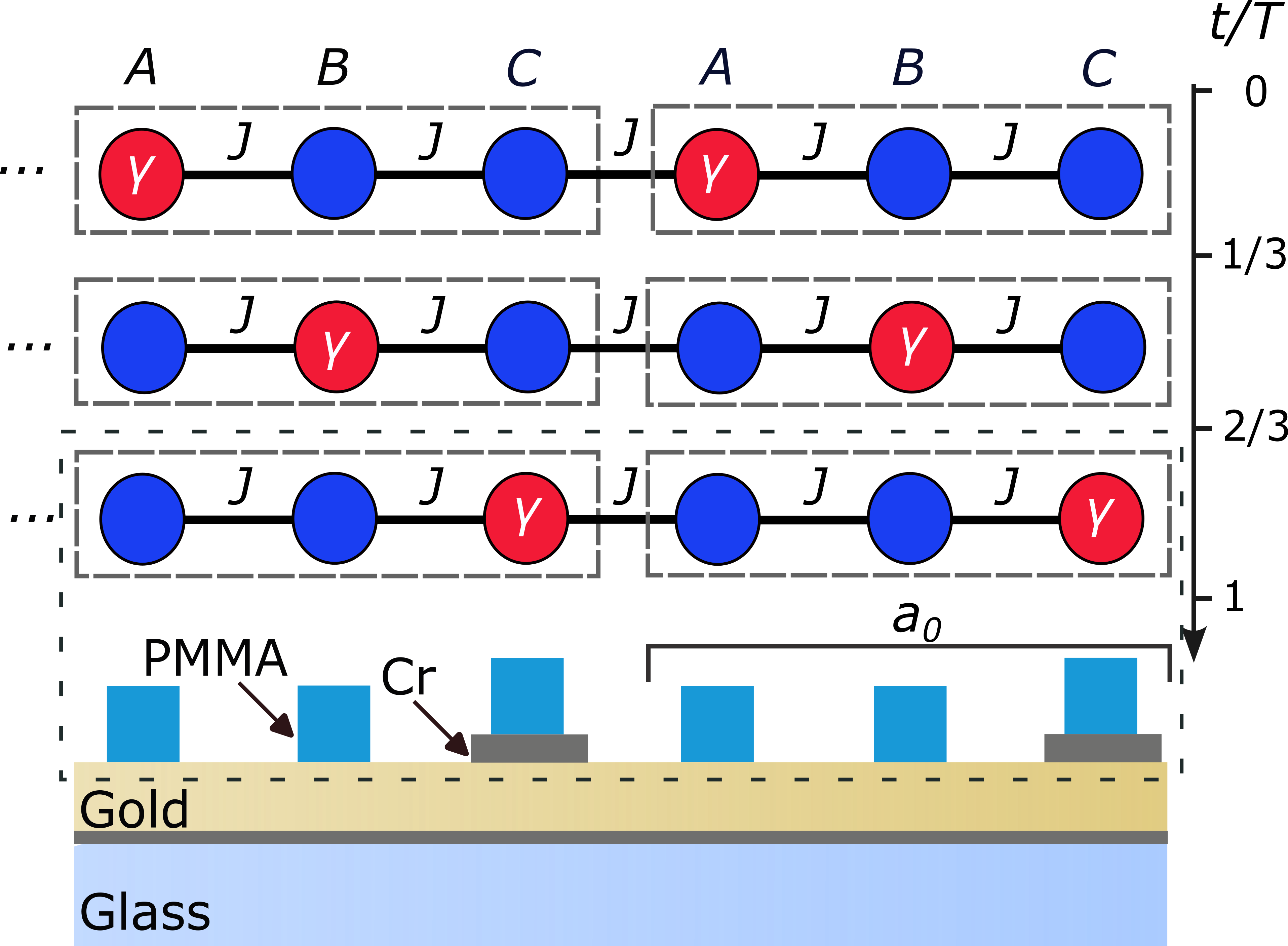}\vspace*{-.2cm}
    \caption{Top - Schematic of the lattice with periodically driven damping for three equidistant times during a driving cycle of period $T$. Lossy sites with dissipation rate $\gamma$ are depicted in red, while non-lossy sites are depicted in blue. $J$ is the hopping constant. Bottom - Sketch of waveguide array corresponding to the last time step of the lattice above. \vspace*{-.6cm}}
    \label{fig:ratchet_sketch}
\end{figure}
The underlying Hamiltonian is proposed as 
a simple tight-binding chain $H_{tb}$ with uniform hopping $J$ and time-dependent dissipation $\mathrm{i}\Gamma(t)$
\begin{align}
    H=H_{tb} -\mathrm{i}\Gamma(t).
    \label{eq: H}
\end{align}
The time periodic damping 
\begin{equation}
    \Gamma  (t) = \sum_j (\gamma_a  (t) a_j^\dagger a_j+\gamma_b  (t) b_j^\dagger b_j+\gamma_c  (t) c_j^\dagger c_j).
\end{equation}
 acts on one site at a time and travels from sites $A$ to $B$ to $C$ through the chain in time-steps of $T/3$ as sketched by the red dots in Fig.~\ref{fig:ratchet_sketch}, 
 where $a_j^\dagger,b_j^\dagger,c_j^\dagger$ ($a_j,b_j,c_j$) are creation (annihilation) operators at site $A,B,C$ of the three-site unit cell $j$ with size $a_0$.
 Mathematically, it is described in terms
of Heavyside functions $\theta$  
\begin{align}
    \gamma_\alpha  (t)= \gamma \ \theta(-1\!-\!2\cos(\omega t\!+\!\varphi_\alpha)) +\gamma_0
    \label{eq:gamma}
\end{align}
where  $\varphi_A=\frac{2\pi}{3}$,  $\varphi_B=0$, $\varphi_C=-\frac{2\pi}{3}$ and $\gamma_0$ is a small background loss, which we will neglect for now. 
The static tight-binding chain $H_{tb}$  has uniform hopping $J$ 
\begin{align}
    H_{tb} = J\sum_j(a_j^\dagger b_j +b_j^\dagger c_j + c_j^\dagger a_{j+1} + h.c.).
\end{align}

{\it Theory.}
The goal is to calculate the propagation of an initial excitation in unit cell $j$ over multiple driving periods through the chain and its losses. 
Due to the step-like nature of the drive in Fig.~\ref{fig:ratchet_sketch} the time-evolution operator for one period $T\!=\!\frac{2\pi}{\omega}$ can be decomposed into a 
product of the three time-steps
\begin{align}
    U(T,0)&= \mathrm{e}^{-\mathrm{i}H\left(\frac{2T}{3}\right)\frac{T}3}\mathrm{e}^{-\mathrm{i}H\left(\frac{T}{3}\right)\frac{T}{3}}\mathrm{e}^{-\mathrm{i}H\left(0\right)\frac{T}{3}}\!,
    \label{eq:U}
\end{align}
where  $U(nT,0)\!=\!U(T,0)^n$. 
We start by considering the case of very strong damping $\gamma\!\rightarrow\! \infty$, 
which forces the particle to move only on lossless sites.  
As shown in appendix A
it is straightforward to evaluate Eq.~(\ref{eq:U}) in order to calculate the probability for a particle to travel exactly one unit cell starting from a site $B$  over one period 
\begin{align}
    |\langle B,j+1|U(T,0)|B,j\rangle |^2 = \left|\sin^3\!\left(TJ/3\right)\right|^2\!.
    \label{eq:PBB}
\end{align}
The probabilities for particles initially at sites $A$ or $C$ to move one complete unit cell are equal to zero. In principle, transport between sites $C\to A,B$ and $B\to A$ is also possible from the off-diagonal elements of  $U(T,0)$, but with probabilities less than 1.  
Hence, perfect transport will occur at resonant frequencies
\begin{align}
    \omega=\tfrac{4}{6m+3}\,J,
    \label{eq:res_con}
\end{align}
where $m\in \mathbb{N}_0$ is the resonance order. 
\begin{figure}  
    \centering
    \includegraphics[width=.9\linewidth]{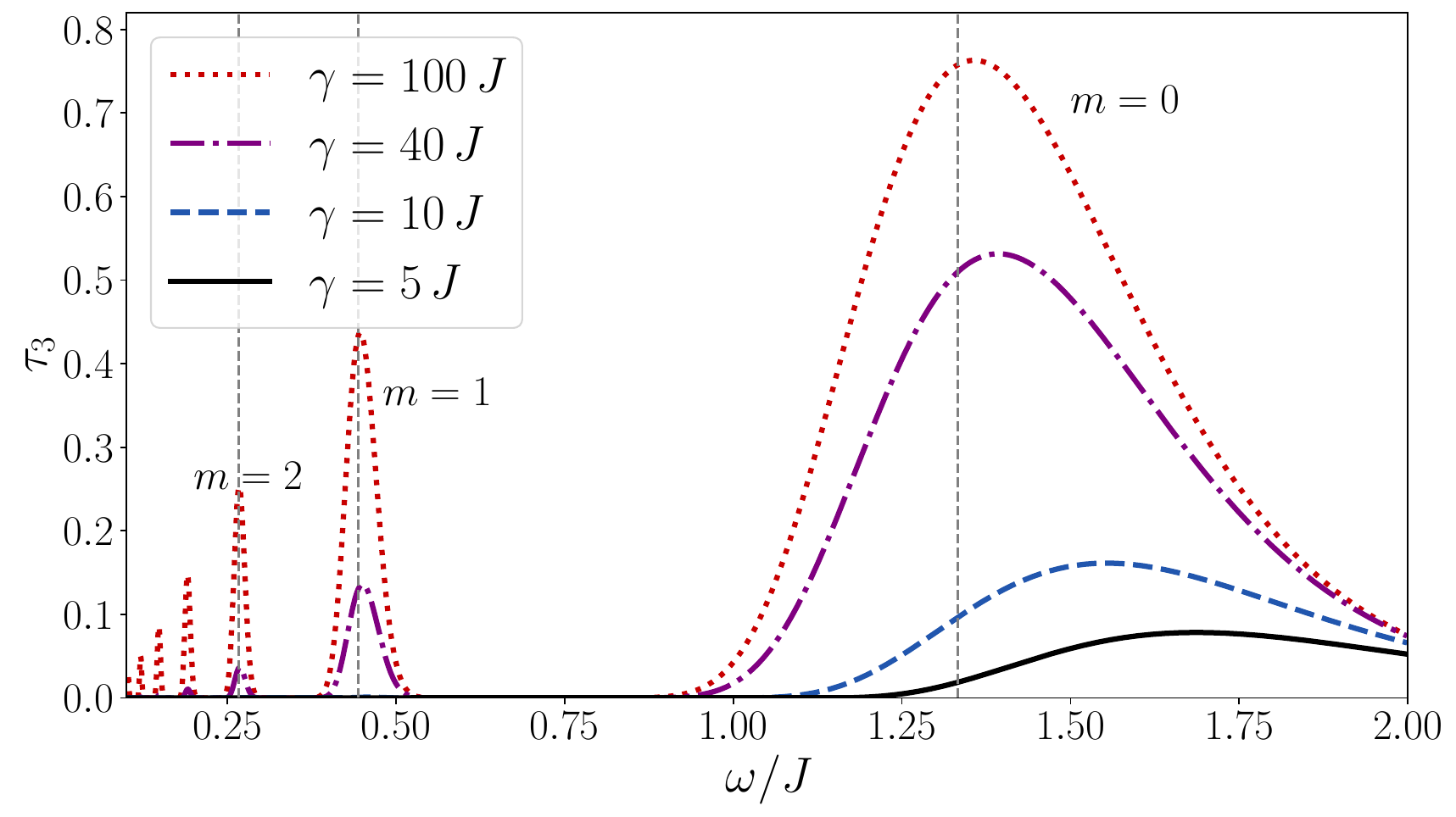} \vspace*{-.5cm}
    \caption{Frequency-dependent transmission of the ratchet as defined in Eq. \eqref{eq:tau} for different losses $\gamma$ calculated after propagation time of 3T. The vertical lines indicate the first three resonance frequencies for the infinite damping case in Eq.~\eqref{eq:res_con}. }\vspace*{-.55cm}
   \label{fig:t}
\end{figure}

We now turn to finite values of $\gamma$ and numerically calculate the transmission over $n$ periods  
\begin{align}
    \uptau_n :=|\langle B,j+n|U(nT,0)|B,j\rangle |^2,
    \label{eq:tau}
\end{align}
as shown in Fig.~\ref{fig:t} for different $\gamma$ and $\!n=\!3$ driving periods.  
Contrary to intuition, the maximum transmission {\it increases} for stronger damping $\gamma$, while the corresponding resonance frequency is approached from above. 
\begin{figure*}
    \centering
    \includegraphics[width=1.8\columnwidth]{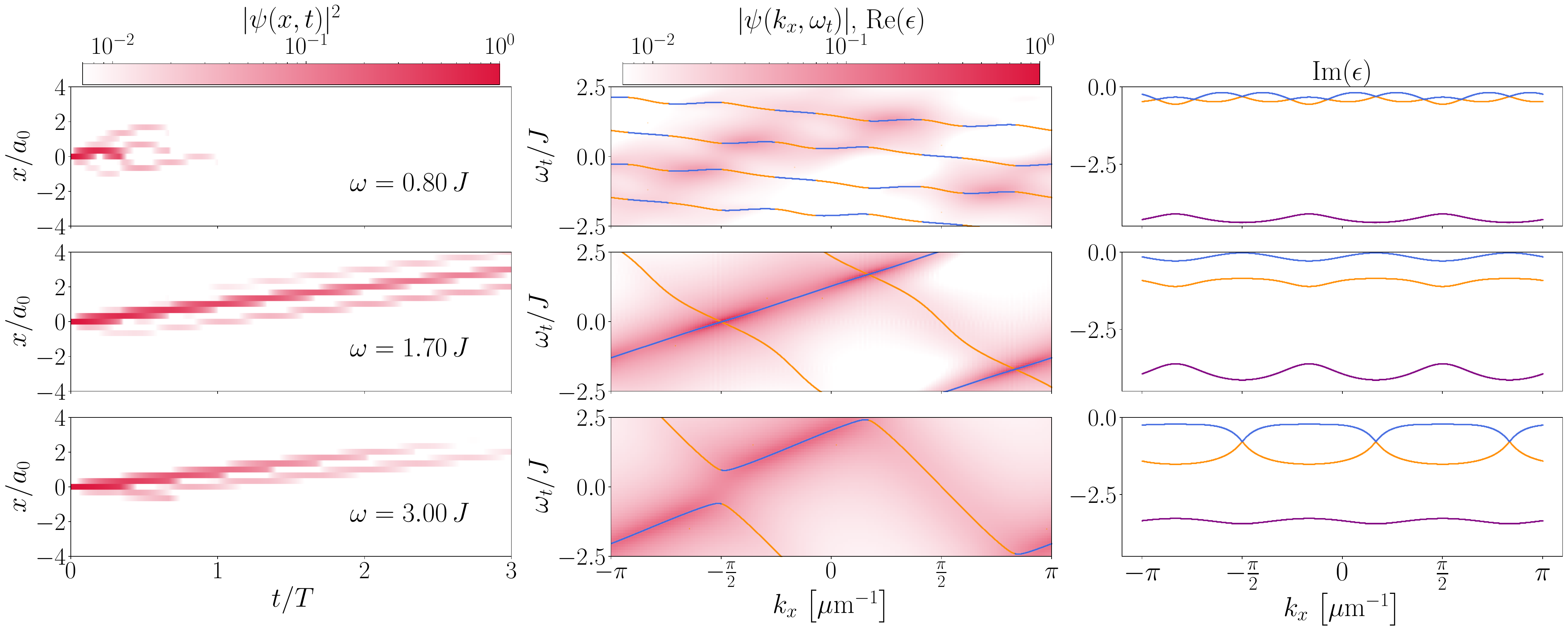} \vspace{-.3cm}
    \caption{  Simulated temporal evolution of $\vert\psi(x,t)\vert^2$ (left) and the Fourier transform in Eq.~(\ref{fourier})  $\vert\psi(k_x,\omega_t)\vert$ (middle) using $\gamma\!=\!5\,J$, $a_0\!=\!3\upmu$m, length $L\!=\!61 a_0$, $t_{max}\!=\!30T$, and an initial single-site excitation $|B,j\!=\!0\rangle$. The lines show the numerically calculated quasienergies separated in real part (middle) and imaginary part (right). \vspace*{-.6cm}}
    \label{fig:sim}
\end{figure*}
For comparison with experiment at different frequencies, we now simulate the system in Eq.~(\ref{eq: H}) over multiple periods considering an 
experimentally feasible dissipation rate of $\gamma\! =\! 5J$. 
Starting with a particle on one site $|B,j\!=\!0\rangle$, the left column of Fig.~\ref{fig:sim} depicts the evolution of the probability density $\vert\psi(x,t)\vert^2$ for three driving frequencies, where  $\ell$ indexes the individual sites at $x\!=\!\ell a_0/3$. 
While a ratchet effect 
is always observed, the transport along $x$ is most efficient for $\omega\!\approx\!1.7J$ corresponding to the maximum for $\gamma\!=\!5J$ in Fig.~\ref{fig:t}. 

The corresponding Fourier transform  
\begin{align} \label{fourier}
    \psi(k_x,\omega_t)=\sum_\ell\mathrm{e}^{-\im k_x x}\int_{0}^T\mathrm{d}t \e^{\im \omega_t t}\psi(x,t) .
\end{align}
is shown in the middle column of Fig.~\ref{fig:sim}.  Near $\omega=1.7J$ the maximum of $|\psi(k_x,\omega_t)|$ follows a straight line in frequency-momentum-space,  which is explained by considering the quasi-energy dispersion in the following.


The  Hamiltonian in Eq.~\eqref{eq: H} can be expressed in momentum space using the Fourier transforms  $\vec{v}_k\!=\!(a_k,b_k,c_k)^T$, where
$\alpha_k\! =\! \frac{1}{\sqrt{N}}\sum_j \mathrm{e}^{\mathrm{-ikj}}\alpha_j, \  \alpha\!=\!a,b,c$  
\begin{align}
    H & =\sum_k \vec{v}_k^\dagger\left[H_k-i\Gamma_k(t)\right]\vec{v}_k \\
  {\rm with} \ \ \ \ \ \ \   \ \ H_k &= \left(\begin{array}{ccc}
0& J & J\mathrm{e}^{-\mathrm{i}k}\\
J &0 & J\\
J\mathrm{e}^{\mathrm{i}k} & J &0
\end{array}\right),\\
    \Gamma_k(t) &= \left(\begin{array}{ccc}
\gamma_a & 0 & 0\\
0 & \gamma_b & 0\\
0 & 0 & \gamma_c
\end{array}\right).
\end{align}
Note that for the $k$ representation we sum over the unit-cell index $j$, contrary to Eq.~\eqref{fourier}, where the sum runs over individual sites. Thus, one Brillouin zone in $k_x$ corresponds to an extended zone scheme of three Brillouin zones in $k\!=\!a_0k_x$. Furthermore, since
the Hamiltonian is periodic in time, the solutions to the Schrödinger equation can be written as Floquet states \cite{shirley1965solution}
\begin{align}
    |\Psi_\alpha (t)\rangle &= \mathrm{e}^{-\mathrm{i}\epsilon_\alpha t}|u_\alpha(t)\rangle,
\end{align}
where $|u_\alpha(t)\rangle$ is time periodic and the multi-index $\alpha$ specifies $k$, band number and Floquet-Brillouin zone. Due to $H$ being non-Hermitian, the quasienergies $\epsilon_\alpha$ can be complex, which can be calculated from the eigenvalues of the time-evolution operator in Eq.~(\ref{eq:U}) over one period \begin{align}
    U(T,0)|\Psi_\alpha (0)\rangle=\mathrm{e}^{-\mathrm{i}\epsilon_\alpha T}|\Psi_\alpha (0)\rangle.
\end{align}

Let us first consider the imaginary parts of the quasienergies in the right panel of Fig.~\ref{fig:sim}, which are always negative, i.e.~all modes are damped, but with much different magnitudes. The most strongly damped mode (purple) cannot contribute to the transport and is neglected in the following. The other two modes (blue and yellow) show smaller imaginary parts, but close to the maximum for $\gamma\!=\!5J$ in Fig.~\ref{fig:t} at $\omega\! = \! 1.7J$ a remarkable behavior develops:  The imaginary part for the lowest damped mode (blue) almost vanishes completely in parts of $k$-space and at the same time the real part shows an almost perfect linear dispersion
independent of $k$.  Accordingly, in the simulation  (left panel of  Fig.~\ref{fig:sim}) we observe very efficient directed transport of $\vert\psi(x,t)\vert^2$ at a specific velocity $v_0\!=\!a_0/T$.  The corresponding Fourier
transform $\vert\psi(k_x,\omega_t)\vert$ shows  pronounces maxima on top of the linear dispersion (middle) with modulations that match the minima in the imaginary part.  For lower and higher frequencies, directed transport is also possible in Fig.~\ref{fig:sim}, albeit over shorter distances, in agreement with
the larger imaginary parts and different velocities.   

  Since equivalent Floquet solutions (replicas) exist for quasienergies shifted by multiples of $\omega$, only
selected branches of the dispersion are shown in Fig.~\ref{fig:sim} (middle). 
An interesting aspect of the quasienergy structure is the appearance of exceptional points, which are defined by degenerate complex energies, where only one eigenstate exists. 
Such singular points have interesting topological properties \cite{PhysRevLett.106.213901,Peng2014,Doppler2016,Hodaei2017,Parto2021,wetter2023observation} and have been observed and studied in the context of dissipative photonic transport \cite{li2023ratchet,mao2025chip}.  A more quantitative 
analysis is given 
in appendix B, where it is shown that exceptional points mark the transition as a function of driving frequency from regions of avoided level crossing in the real spectrum to regions with Dirac dispersion and possible resonances.  

Finally, we want to comment a uniform damping $\gamma_0$ in Eq.~(\ref{eq:gamma}), which corresponds to adding a constant on the diagonal of the Hamiltonian \eqref{eq: H}.  This results in a change $\epsilon\!\rightarrow \!\epsilon \!-\!\im \gamma_0$, so the damping of all modes is increased, but the spectrum remains the same otherwise.

\begin{figure}[b]
    \centering  \vspace*{-.6cm}
    \includegraphics[width=0.45\textwidth]{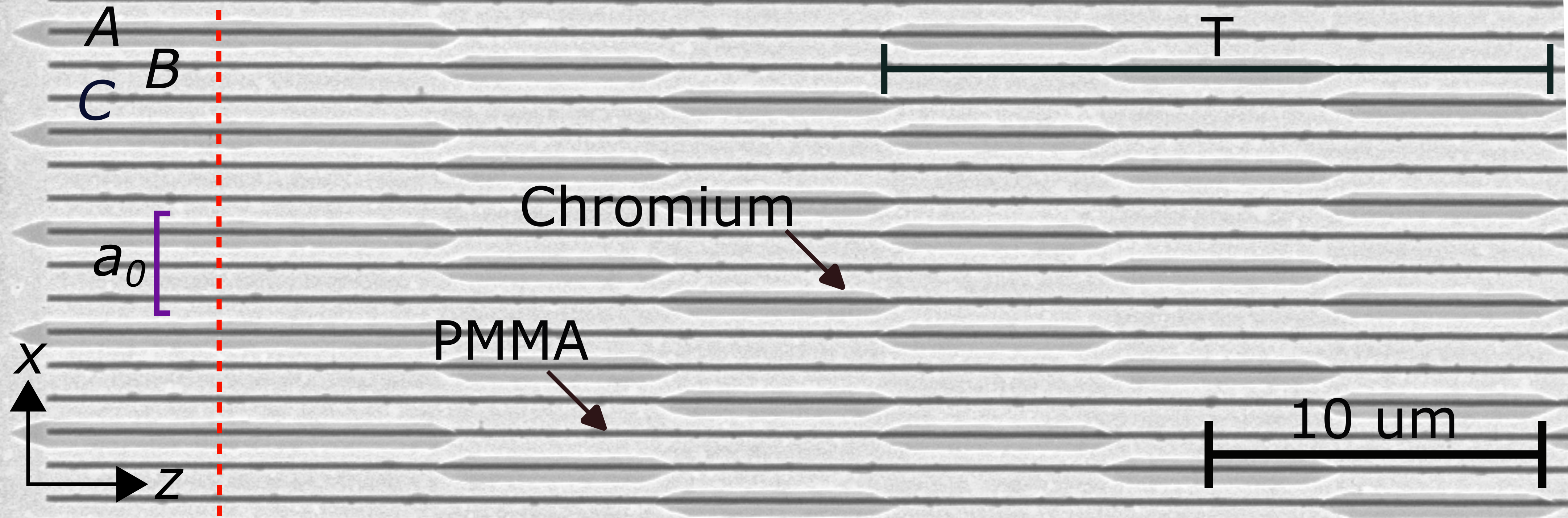} \vspace*{-.2cm}
    \caption{Scanning electron micrograph of the ratchet realization in a plasmonic waveguide array. The red dashed line marks the boundary of the excitation region, $a_0$ - the size of the unit cell, $A$, $B$, $C$ indicate sites within the unit cell - possible inputs, $T$ is the period of modulation.  }
    \label{fig:sample}
\end{figure}

{\it Experiments.}
We realize a dissipation-engineered ratchet using arrays of coupled dielectric-loaded surface plasmon polariton waveguides (DLSPPWs). The mathematical analogy between the tight-binding Schrödinger equation and the Helmholtz paraxial equation enables us to directly map time to propagation distance \cite{christodoulides2003discretizing, longhi2009quantum} and implement time modulation \cite{Fedorova,sidorenko2022real}. The array consists of waveguides at equal distance $d\!=\!1\mathrm{\upmu m}$ that translate into constant couplings  $ J \!=\! 0.089\mathrm{\upmu m}^{-1}$ (measured in an auxiliary experiment with two waveguides). Local dynamic dissipation is achieved by depositing chromium patches beneath the waveguide \cite{FedorovaDirFilter} in a periodic pattern along the $x$ and $z$ (time) axes (see Fig.~\ref{fig:sample}), resulting in dynamic dissipation rates $ \gamma\! =\! 0,~1.96J,~4.88J$ (evaluated using finite element calculations) corresponding to no Cr, and the Cr thicknesses of 8 and 20 nm, respectively. The constant loss in a single waveguide is $\gamma_0 \!=\! 0.185J$. Note that such homogeneous losses only cause overall exponential decay of the probability density in real space and spectral line broadening in Fourier space, but otherwise do not alter the system dynamics.

\begin{figure*}
    \centering
    \includegraphics[width=1.4\columnwidth]{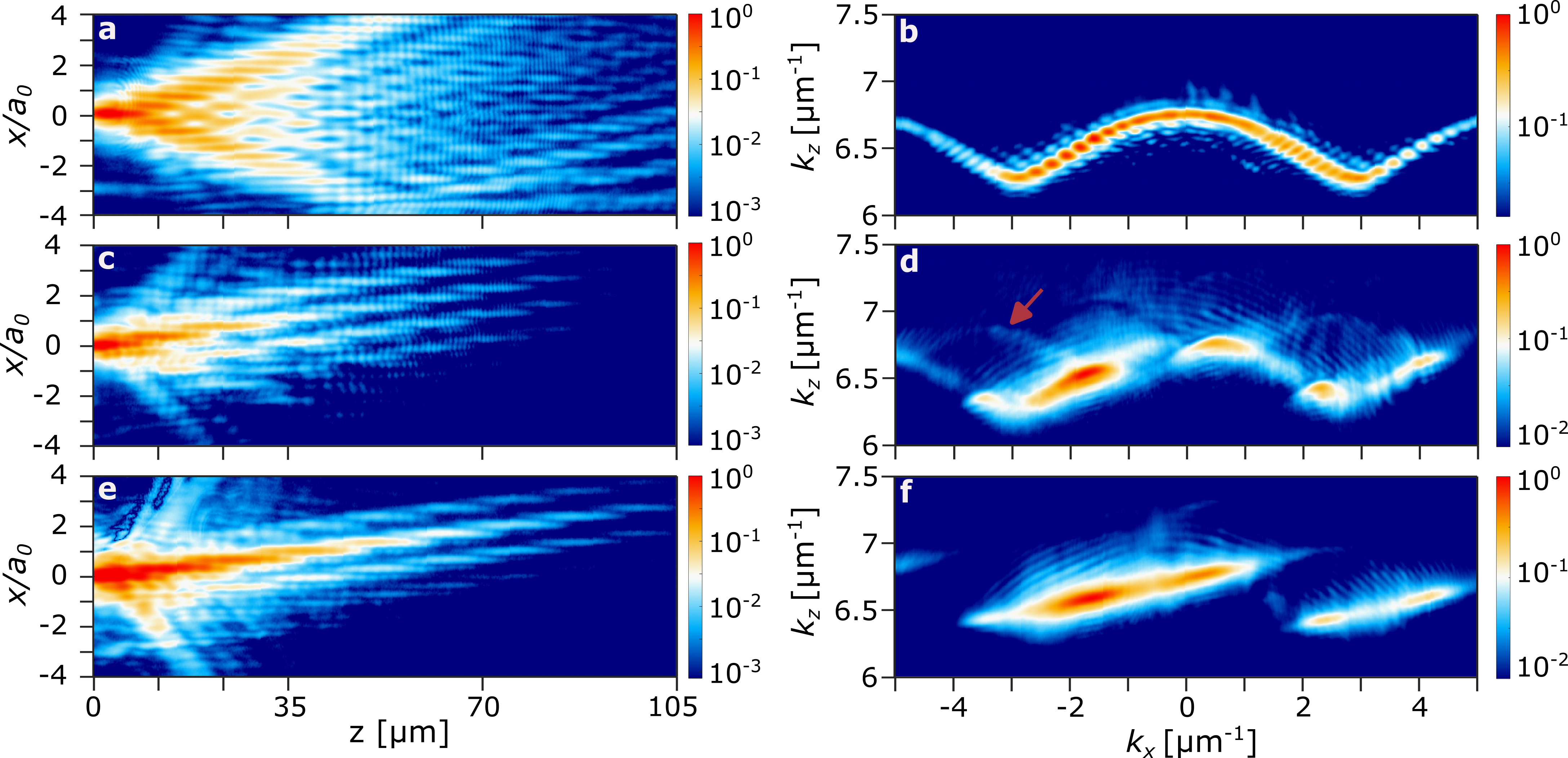} \vspace*{-.2cm}
    \caption{Measured real-space (left) and Fourier-space (right) SPP intensity distributions for different dissipation rates and single-site excitation at site $B$ for driving period $P\!=\!35\upmu$m corresponding frequency $\omega\!=\!2.02J$ (a) and (b) show results for no loss, (c) and (d) - $\gamma\!=\! 1.96J$, where arrow highlights Floquet replica, (e) and (f) - $\gamma\!=\! 4.88J$.\vspace*{-.3cm}}
    \label{fig:exp_data_low_loss}
\end{figure*}
\begin{figure*}
    \centering
    \includegraphics[width=1.4\columnwidth]{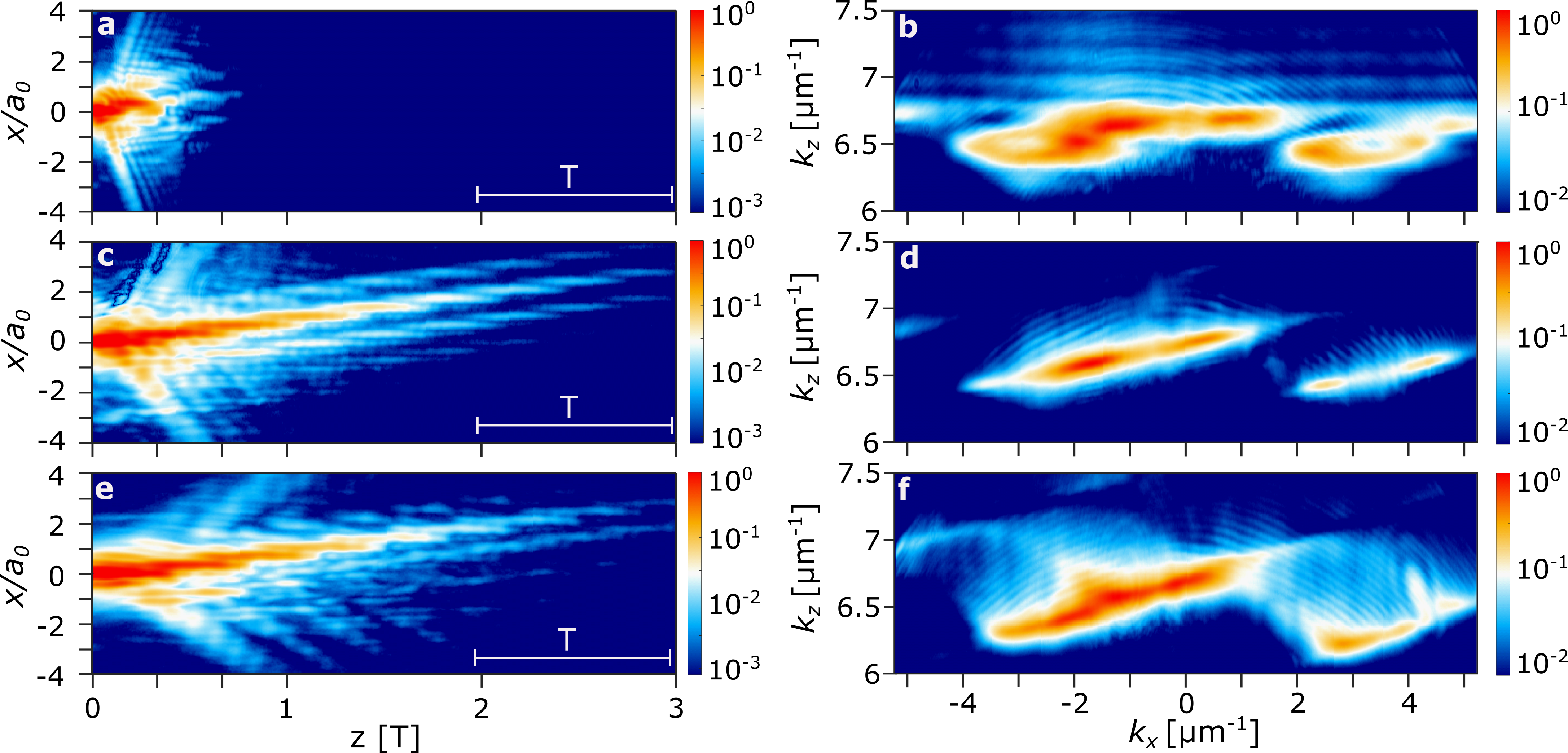} \vspace{-.3cm}
    \caption{Measured real-space (left) and Fourier-space (right) SPP intensity distributions for different driving frequencies outside and within resonance order $m=0$ and single-site excitation at site $B$ for dissipation of $\gamma\!=\! 4.88J$ (a) and (b)  off resonant regime ($P\!=\!90\upmu$m corresponding frequency $\omega\!=\!0.78J$), (c) and (d) show results for - intermediate frequency regime ($P\!=\!35\upmu$m corresponding frequency $\omega\!=\!2.02J$), (e) and (f) - high frequency regime ($P\!=\!20\upmu$m corresponding frequency $\omega\!=\!3.53J$).\vspace*{-.6cm}}
    \label{fig:exp_data}
\end{figure*}

SPPs are excited by focusing a TM-polarized laser beam with $\lambda_0\!=\!980$~nm onto the grating coupler deposited on top of the chosen waveguide for a single-site excitation. SPP transport within the array was monitored through real- and Fourier-space leakage radiation microscopy. The measured SPP intensity in real space $I(x,z)$ corresponds to the probability density $\vert\psi(x,t)\vert^2$ in the theory simulations above~\cite{bleckmann2017spectral} and we also consider the SPP intensity in Fourier space $I(k_x,k_z)$. 
The measurement method is described in appendix D.

From the numerical simulations in Fig.~\ref{fig:t}, we expect that both the driving frequency and the dissipation strength influence the transmission maximum. First, we explore how increased losses give rise to the ratchet effect. Figure~\ref{fig:exp_data_low_loss} shows the measured real space and Fourier space SPP intensity distributions for constant frequency $\omega\!=\!2.02J$ corresponding to driving period $T\!=\!35\upmu$m, and dissipation rates of 
$\gamma\! =\! 0,~1.96J,~4.88J$. The upper row displays the case of $\gamma\!=\!0$ (Fig.~\ref{fig:exp_data_low_loss}a,b), resulting in an array of identical equidistant waveguides without time-periodic modulation. The real-space image features ballistic expansion of the wave packet with no preferred direction in addition to a characteristic interference pattern \cite{bleckmann2017spectral}. The corresponding Fourier-space distribution comprises a cosine-shaped band. The following row in Fig.~\ref{fig:exp_data_low_loss}c,d represents the low-loss case of $\gamma\! =\!1.96J$, which results in a slight confinement of the wave packet and its directed displacement over time. In Fourier space we observe a slight deformation of energy bands as well as the appearance of Floquet replicas, i.e., copies of the band structure shifted by multiples of the driving frequency, marked by the arrow. Additionally, we find these bands unevenly populated. This is caused by the initial conditions together with a loss of symmetry in the Hamiltonian, as well as the emergence of damping depending on $k$. Further increasing dissipation to $\gamma=4.88J$ (Fig.~\ref{fig:exp_data_low_loss}e,f) leads to clear directional transport of SPPs within the lattice. 
The corresponding Fourier-space SPP intensity distribution differs drastically from the preceding regime; it consists of linear bands, with a predominant population of positive velocities only.
Notably, increasing losses lead to better directional transport.

Next, we perform a frequency sweep for a fixed dissipation rate of $ \gamma\! =\!4.88J$. First, we start with the off-resonant frequency $\omega\!=\!0.78J$. 
In this regime (Fig.~\ref{fig:exp_data}a,b), we observe strongly suppressed transport, and the corresponding Fourier space image displays no clear structure. In the intermediate-frequency regime near resonance $\omega\!=\!2.02J$ in Fig.~\ref{fig:exp_data}e,f, the real-space measurement displays a wave packet moving one unit cell per driving period upward. The main feature of this momentum-resolved spectrum is linear bands with a constant positive slope corresponding to upwards-moving states. There are distinct higher-intensity spots at specific $k$ values highlighting oscillations in the imaginary part of the quasienergies. 
In the case of high frequency $\omega\!=\!3.53J$ in Fig.~\ref{fig:exp_data}g,h, the real-space SPP intensity distribution shows a wave packet moving upwards. The corresponding momentum-resolved spectrum show dominantly populated bands with positive slope, together with some weaker excitation of bands with negative velocity.


{\it Conclusions.}
To conclude, we propose and implement a Floquet-dissipative ratchet design that enables control over the directionality of transport solely through time-periodic dissipation. Experimentally the ratchet was realized using evanescently coupled waveguides with periodic losses. From real- and Fourier space measurements, we show strong nonreciprocal transport at all quasimomenta for a sufficient dissipation rate. Counter\-intuitively, the transmission is increased as the losses increase, marking strong time-tailored dissipation as a feature instead of a disadvantage. Using Floquet theory, we determine the complex quasienergy band structure of the non-Hermitian Hamiltonian and identify resonant frequency regimes of strong unidirectional transport. The main feature is the population of a linear band with minimal, near-constant losses, which was confirmed experimentally and in numerical simulations.  Resonant regimes with linear Dirac dispersion are separated from avoided level crossing
regimes by sharp transitions with characteristic exceptional points in the spectrum.

Achieving rectification solely by time-periodic losses with non-Hermitian Hamiltonian time evolution 
distinguishes our Floquet-dissipative ratchet design from both Brownian motors and 
Hermitian ratchets. It provides an expansion to the toolbox of quantum functionalities and furthers understanding of using dissipation in open quantum 
systems control~\cite{PhysRevB.106.L180302,marquet2024harnessing,zhang2025observation,bk7q-6r9d,PhysRevApplied.20.044019}. 

We are thankful for discussions with C.~Dauer and S.~J\"ager.  Support from the DFG within SFB/TR 185 (277625399) is acknowledged.


\bibliography{bibliography}

\appendix
\section{Appendix A: Limit of large damping $\gamma$}
\label{app:largeGamma}
Using Eq.~\eqref{eq:U} we can calculate the time evolution operator over one period $U(T,0)$. 
Let us consider one unit cell with periodic boundary conditions. This already provides great insights into the transport processes at large $\gamma$, since the damping effectively isolates a unit cell at each timestep.
We find 
\begin{align*}
\mathrm{e}^{-\mathrm{i}H\left(0\right)\frac{T}{3}}&=\exp\left[-\mathrm{i}\left(\begin{array}{ccc}
-\mathrm{i}\gamma & 1 & 1\\
1 &0 & 1\\
1 & 1 & 0
\end{array}\right)\frac{JT}{3}\right]\\
&= S\left(\begin{array}{ccc}
\mathrm{e}^{-\mathrm{i}\lambda_0\frac{JT}{3}} &0&0\\
0&\mathrm{e}^{-\mathrm{i}\lambda_1\frac{JT}{3}}&0\\
0&0&\mathrm{e}^{-\mathrm{i}\lambda_2\frac{JT}{3}}
\end{array}\right)S^{-1},
\end{align*}
where $S$ is the corresponding transformation matrix and
\begin{align*}
\lambda_0&=-1,\\
\lambda_1&=\frac{1}{2}(1-\sqrt{9+2\mathrm{i}\gamma-\gamma^2}-\mathrm{i}\gamma ) \rightarrow-\im\gamma,\\
\lambda_2&=\frac{1}{2}(1+\sqrt{9+2\mathrm{i}\gamma-\gamma^2}-\mathrm{i}\gamma )\rightarrow 1
\end{align*}
are the eigenvalues of the Hamiltonian. The arrow indicates the limit of large $\gamma$. Analogously, we can calculate the other exponentials in 
Eq.~\eqref{eq:U}. We find
\begin{align*}
    &\lim_{\gamma \rightarrow \infty}U(T,0)\\
    =& \left(\begin{array}{ccc}
0&-\cos\left(J\frac{T}{3}\right)\sin^2\!\left(J\frac{T}{3}\right) & -\mathrm{i}\cos^2\!\left(J\frac{T}{3}\right)\sin\left(J\frac{T}{3}\right)\\
0& -\mathrm{i}\sin^3\!\left(J\frac{T}{3}\right) & -\cos\left(J\frac{T}{3}\right)\sin^2\!\left(J\frac{T}{3}\right) \\
0&0&0
\end{array}\right)\!.
\end{align*}
The matrix elements $U(T,0)_{(i,j)}$ give the probability to find an excitation at site $j$ at site $i$ after one driving period. Thus we obtain Eq. \eqref{eq:PBB}.

\section{Appendix B: Exceptional points}\label{app:QE}
\begin{figure}
    \centering
    \subfigure{\includegraphics[width=0.95\linewidth]{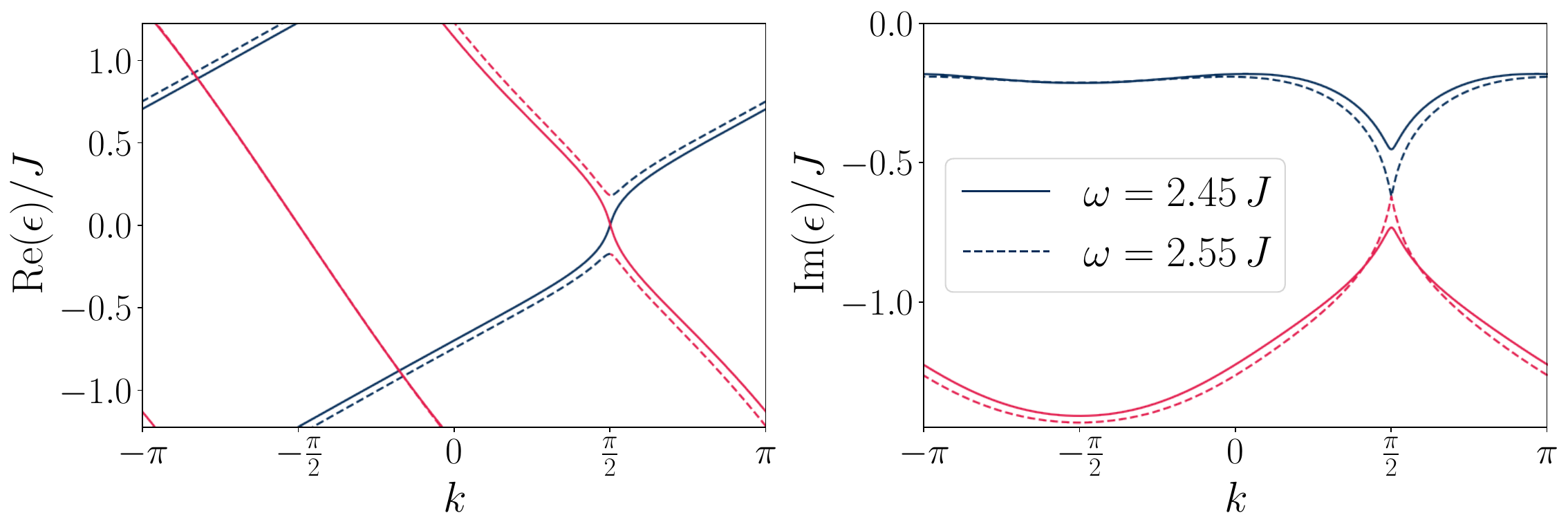}}
    \subfigure{\includegraphics[width=0.95\linewidth]{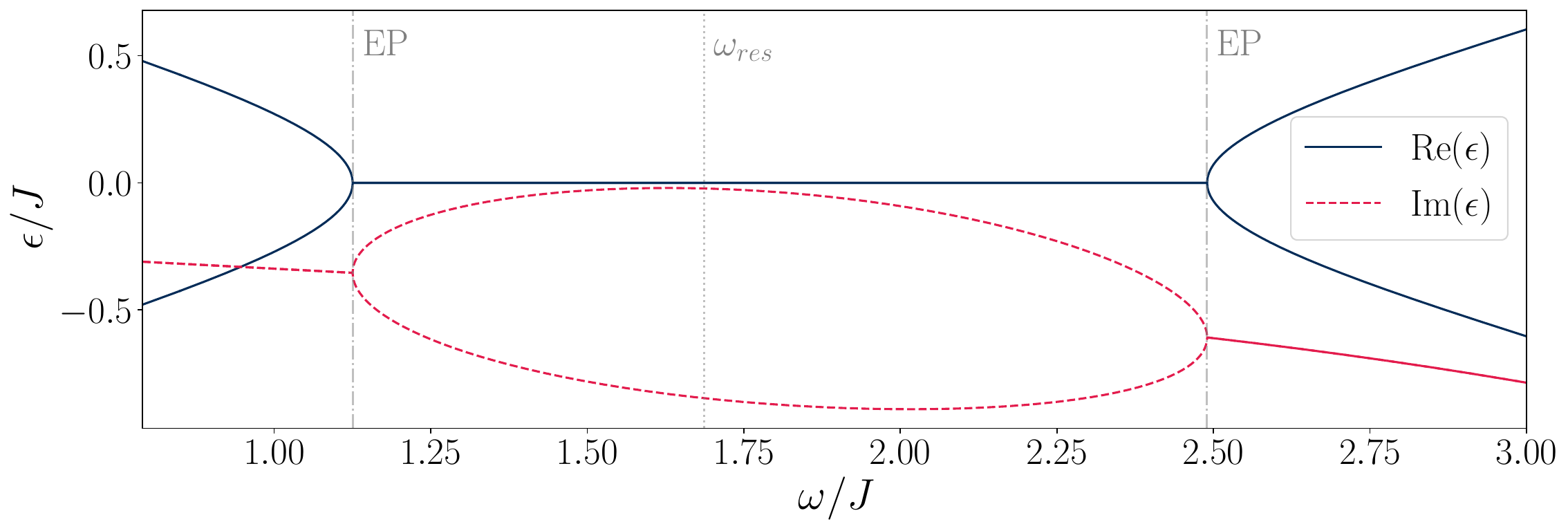}}
    \caption{Quasi-energies in the vicinity of an exceptional point at $\gamma=5\,J$. The upper plot shows the real and imaginary parts of the quasienergy for selected frequencies. The lower plot shows the quasienergies at $k\!=\!\frac{\pi}{2}$ as a function of frequency. \vspace*{-.6cm}}
    \label{fig:excaptionalPoint}
\end{figure}

One notable aspect in Fig.~\ref{fig:sim} is the observation of level crossings, which appear {\it either} in the imaginary part ($\omega\!=\!0.8J, \ 3J$) {\it or} in the real part ($\omega \!= \!1.7J$).   A more qualitative 
analysis between those region is done for $\gamma\!=\!5J$ in Fig.~\ref{fig:excaptionalPoint}(top) as a function of $k$ and $\omega$.  A sharp transition is observed at $k\!=\!\pi/2$ between gaps in the real and
the imaginary parts respectively, occurring at an exceptional point (EP) \cite{Heiss_2012,RevModPhys.93.015005}.  The transition of the spectrum is demonstrated in  Fig.~\ref{fig:excaptionalPoint}(bottom),
where the middle region with Dirac spectrum can host a resonance, separated by two EP at $\omega_{\rm EP} \!\approx \!1.126J, \ 2.489J$. There are two eigenmodes everywhere, except at the 
EP where not only the complex energies are degenerate, but also the two modes coalesce into one single eigenvector, which we have verified in the simulations.


\section{Appendix C: Sample fabrication}\label{appendix D}
The DLSPPW arrays with locally modulated dissipation are fabricated using a two-step electron beam lithography (EBL) process. 
The sample preparation begins with the evaporation of 60 nm of Au and 5 nm of Cr for adhesion on a cleaned glass substrate surface. Then, the sample is spin-coated with a bilayer of polymeric resist poly(methyl methacrylate) (PMMA) of two molecular weights (950K and 495K). In the first EBL step, we utilize PMMA as a positive-tone resist to fabricate a template for the lossy regions and alignment markers. The areas exposed to the electron beam are dissolved in a developer, and the required thickness of Cr is evaporated on top of the substrate. After the lift-off process, we end up with Cr stripes and alignment markers at the predefined positions. The width of each Cr stripe is set to $0.6~\mathrm{\upmu m}$. Every patch is tapered on both sides to minimize edge scattering. Then the sample is again spin-coated with PMMA, and the second EBL step is performed. Now we fabricate the DLSPPW arrays on top of the Cr stripes using the markers for the alignment. In this step, PMMA acts as a negative tone resist, which is achieved by increasing the applied electron dose~\cite{block2014bloch}. Finally, the samples are developed in acetone.
The atomic force microscopy measurements revealed that the applied electron dose results in a mean waveguide height of $130$~nm and a width of $285$~nm, allowing us to work in a single-mode regime at a vacuum wavelength of $\lambda\!=\!0.98\mathrm{\upmu m}$. For these geometric parameters, the propagation constant of the guided mode is $\beta\!=\!\beta'\!-\!i\beta''\!=\!\mathrm{const}$, $\beta'\!=\!6.63\mathrm{\upmu m}^{-1}$ and $\beta''\!=\!0.0165\mathrm{\upmu m}^{-1}$ (obtained by measuring the propagation length of the SPPs). The grating for SPP excitation is deposited only onto the two non-lossy waveguides in the unit cell (inputs $B$ or $C$), while the extension of others to this region is needed to prevent fire-end excitation of the adjacent waveguides. As shown in Fig.~\ref{fig:sample}, the site $A$ in each unit cell contains an extended Cr patch to minimize unwanted excitation and propagation of SPPs.

\begin{figure}
    \centering
    \includegraphics[width=0.38\textwidth]{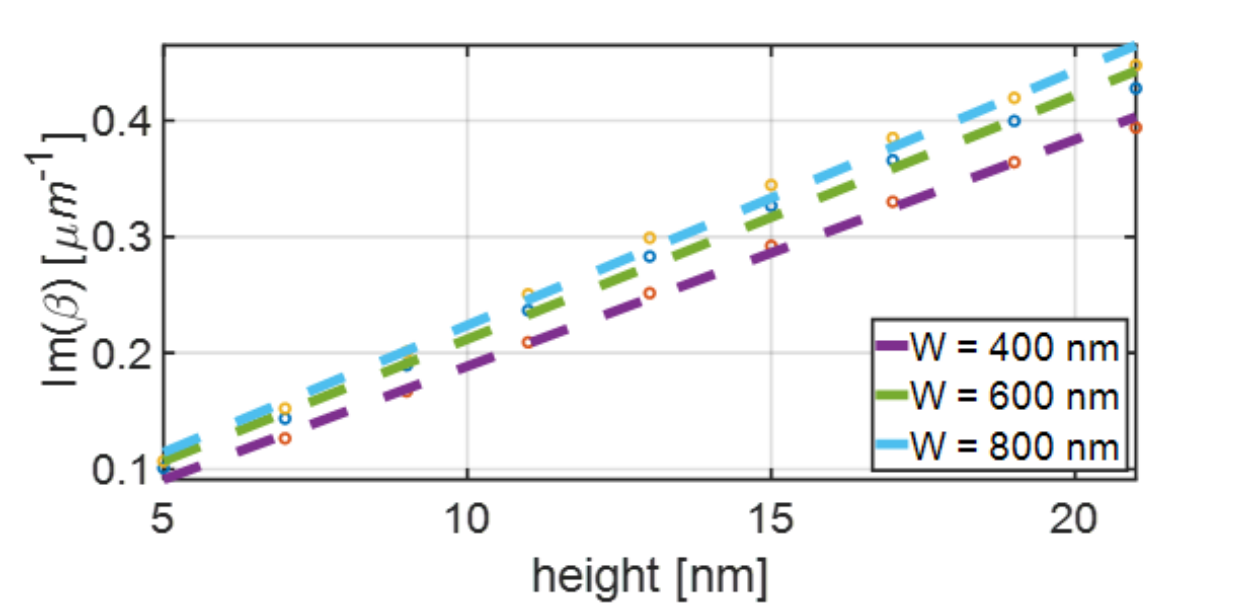}
    \caption{Calculated dissipation in a single waveguide $\Im (\beta)$ versus height of Cr patch for three widths. \vspace*{-.4cm} }
    \label{fig:losses}
\end{figure}

Fig.~\ref{fig:losses} shows the range of the imaginary part of the propagation constant (loss) that can be obtained by varying the geometry of the Cr patch.

\section{Appendix D:  Leakage Radiation Microscopy}\label{appendix E}
SPPs are excited by focusing a TM-polarized laser beam with $\lambda_0$=980~nm  (NA of the focusing objective is 0.4) onto the grating coupler deposited on top of the chosen waveguide. The propagation of SPPs in the array is monitored by real- and Fourier-space leakage radiation microscopy (LRM). The leakage radiation, as well as the transmitted laser beam, are both collected by a high NA oil immersion objective (Nikon 1.4 NA, 60x Plan-Apo). The transmitted laser was filtered out by placing a knife-edge at the intermediate back focal plane (BFP) of the oil immersion objective. The remaining radiation was imaged onto an sCMOS camera (Andor Marana). The SPP intensity distributions in real space were recorded in the real image plane, while the intensity distribution in momentum space was obtained by imaging the BFP of the oil immersion objective.

We note that, in any experiment, in addition to ohmic losses, the propagation of SPPs on the gold substrate and the tailored Cr stripes is inevitably accompanied by additional losses from leakage radiation and imperfections in the metal film. As a result, the bands show additional line broadening in Fourier space and scattering in real space compared to the numerical calculations.

\section{Appendix E: Initial excitation on site $C$}\label{appendix F}
\begin{figure}[b!]
    \subfigure{\includegraphics[width=0.45\linewidth]{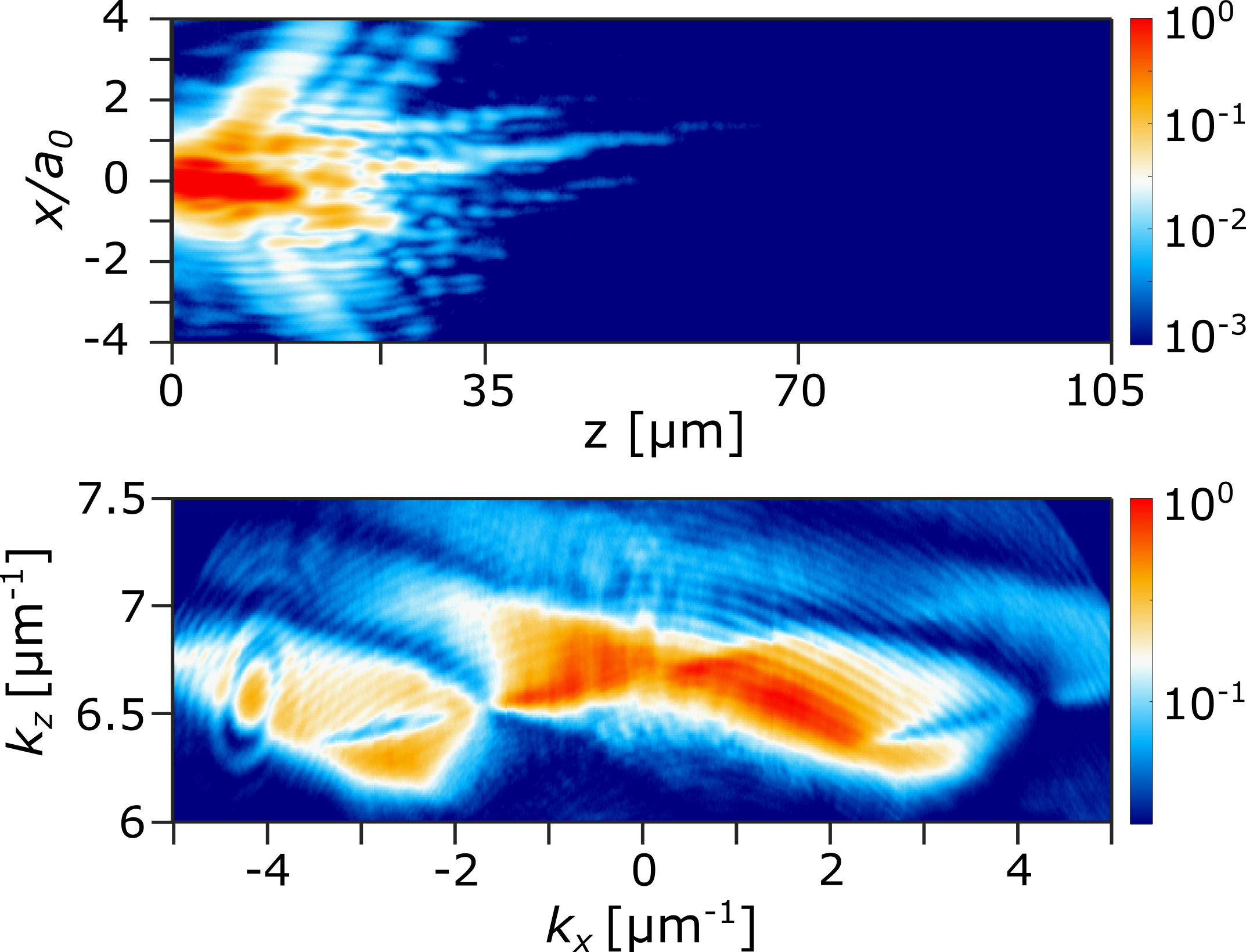}}
    \subfigure{\includegraphics[width=0.45\linewidth]{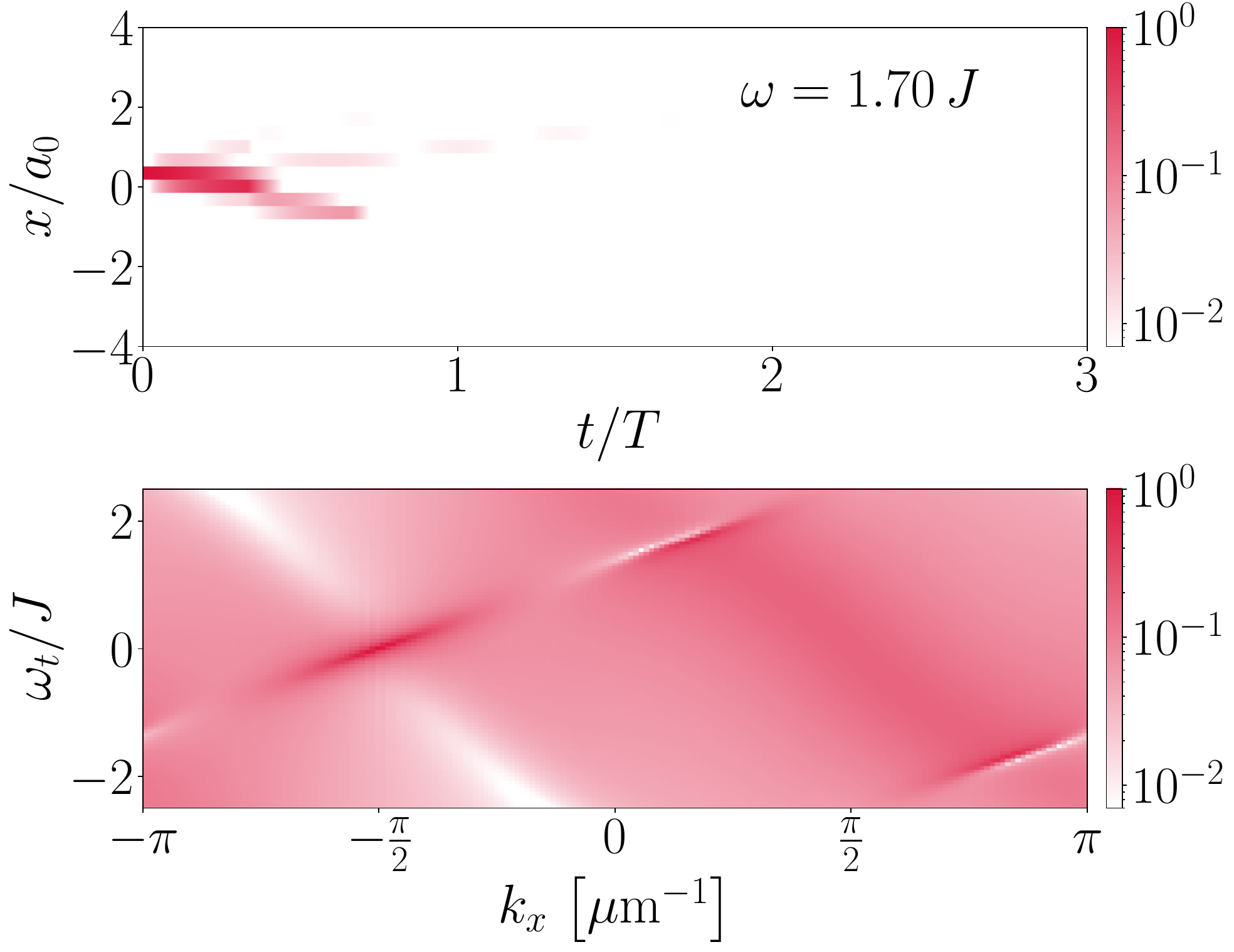}}
    \caption{Left - measured real (top) and Fourier (bottom) leakage radiation microscopy images for the intermediate frequency regime and single-site excitation at site $C$ for dissipation of $\gamma= 4.88J$. Right - the corresponding simulation plots. }
    \label{fig:inp_C}
\end{figure}

To confirm the independence of the initial conditions, we perform simulations as well as measurements for an initial excitation at the site $C$ at resonance. Figure~\ref{fig:inp_C} shows that in this case transport rectification is still observed; however, unfavorable excitation conditions and subsequent coupling into the mode with larger damping lead to most of the intensity being dissipated. The corresponding Fourier-space measurement reveals a significant difference from the site $B$ excitation, characterized by populated, extremely broadened bands with negative slopes that compromise the resolution between the positive and negative movers. The bands with positive-sloped bands can still be distinguished by the accumulation of intensity in a thin line. From this we conclude that, even under unfavorable excitation conditions, the ratchet effect persists, blocking the unwanted transport. This is consistent with the theory, according to which all downward-moving modes are strongly 


\end{document}